\documentclass[preprint,showpacs,preprintnumbers,amsmath,amssymb]{revtex4}

\usepackage{graphicx}
\usepackage{dcolumn}
\usepackage{bm}

\begin{document}

\preprint{}

\title{Metric-Dependent Probabilities that Two Qubits are Separable}

\author{Paul B. Slater}%
\email{slater@kitp.ucsb.edu}
\affiliation{%
ISBER, University of California, Santa Barbara, CA 93106-2150\\
}%
\date{\today}

\begin{abstract}
In a previous study ({\it Quant. Info. Proc.} 1, 397 [2002]), we formulated
a conjecture that arbitrarily coupled qubits --- describable by 
$4 \times 4$ density matrices --- are separable with
an {\it a  priori} probability of $\frac{8}{11 \pi^2} \approx 0.0736881$. For 
this purpose, we employed
the normalized volume element of the {\it Bures} (minimal monotone) metric as
a probability distribution over the {\it 
fifteen}-dimensional
convex set of $4 \times 4$ density matrices. Here, we provide
further/independent (quasi-Monte Carlo numerical integration) 
evidence of a {\it stronger} nature 
(giving an estimate of 
0.0736858 {\it vs.} 0.0737012 previously) for this conjecture. 
Additionally, employing a certain {\it ansatz}, we
 estimate
the probabilities of separability based on certain other monotone metrics of interest. However, we find ourselves, at this point,
unable to convincingly conjecture exact simple 
formulas for these new (smaller) probabilities.
\end{abstract}

\maketitle
An arbitrary state of two quantum bits (qubits) is describable by a $4 \times 4$ density
matrix --- an Hermitian, nonnegative definite matrix having trace unity.
 The convex set of all such density matrices is 15-dimensional in
nature \cite{vanik,fano}. Endowing this set with the statistical
distinguishability (SD) metric \cite{sam}, we were able in \cite{slaterqip} to
address the question (first essentially raised in \cite{ZHSL} and later
studied further in \cite{zycz2,slaterA,slaterC}) 
of what proportion of the 15-dimensional convex
set (now a Riemannian manifold) is separable (classically correlated)
\cite{werner}. The Peres-Horodecki partial transposition criterion 
\cite{asher,michal}  provides
a convenient necessary {\it and} sufficient condition for testing for 
separability in the cases of
qubit-qubit and qubit-qutrit pairs \cite{qq}.

In \cite{slaterqip} we had taken 
the probability of separability of two arbitrarily paired 
qubits to be the ratio of the SD volume ($V_{SD}^{s}$)
occupied by the separable states to the SD volume ($V_{SD}^{s+n}$)
occcupied by the totality of separable {\it and} nonseparable states.
We utilized a conjecture (combining exact and numerical results) that
\begin{equation} 
\label{eq1}
V_{SD}^{s+n} = \frac{\pi^8}{1680} \approx 5.64794,
\end{equation}
the full veracity of which has since been formally established, 
within an impressively broad 
analytical 
framework \cite[eq. (4.12)]{hans}.
 Then, on the basis of certain extended, advanced quasi-Monte Carlo computations
(scrambled Halton sequences \cite{giray1}),
used  for numerical integration in high-dimensional spaces, we had been led to further conjecture in \cite{slaterqip} that
\begin{equation} 
\label{eq2}
V_{SD}^{s} = \frac{\pi^6}{2310} \approx 0.416186.
\end{equation}
(In an earlier study \cite{slaterC}, a number of quite 
surprisingly simple {\it 
exact} results
were obtained using {\it symbolic} integration, for certain specialized 
[low-dimensional] 
two-qubit scenarios --- which had led us to entertain the possibility 
in \cite{slaterqip} of an
exact probability of separability in the full 15-dimensional setting.)
For a scrambled Halton sequence consisting of 65 million points
distributed over a 15-dimensional hypercube, we had 
obtained in \cite{slaterqip} estimates of 5.64851 and .416302
of $V^{s+n}_{SD}$ and $V^s_{SD}$, respectively.

 Let us note that the
 interestingly-factorizable denominators
in (\ref{eq1}) and (\ref{eq2}), that is,
  $1680 = 2^4 \cdot 3 \cdot 5 \cdot 7$ and
$2310 = 2 \cdot 3 \cdot 5 \cdot 7 \cdot 11$ have  been conjectured
elsewhere
\cite[seq. A064377]{sloane}
 (available on-line at
 www.research.att.com/~njas/sequences/Seis.html)
 to be the two {\it greatest} numbers ($k$) for which
the sum of the {\it fourth} powers of the divisors of $k$
 exceeds the {\it fifth} power
of the Euler totient function $\phi(k)$.
($\phi(k)$ is the number of positive integers --- including 1 --- less 
than $k$ and
relatively prime to $k$ \cite{conway}.)

The fundamental 
conjecture (\ref{eq2}), of course, directly
implied the further conjecture that the SD probability
of separability is
\begin{equation} \label{fundamental}
P_{SD}^{s} = \frac{V_{SD}^{s}}{V_{SD}^{s+n}} = \frac{8}{11 \pi^2} \approx 0.0736881.
\end{equation}
Now, since the SD metric is identically {\it four}
 times the Bures (minimal monotone)
metric \cite{sam}, we have the immediate consequence that $P_{Bures}^{s} =
P_{SD}^{s}$. (In regard to arbitrarily coupled qubit-{\it qutrit} pairs,
we reached  in \cite{qq}, by similar methods, the conjecture that the 
corresponding probability of
separability was the ratio ($\approx .00124706$)
 of $2^{20} \cdot 3^3 \cdot 5 \cdot 7$
to the product of $\pi^3$ and the seven 
consecutive prime numbers lying between
19 and 43.)

The Bures metric plays the role of the {\it minimal}
 monotone metric. The monotone
metrics comprise an infinite (nondenumerable) class \cite{petz1,petz2,lesniewski}, generalizing the (classically {\it 
unique}) Fisher information metric \cite{kass}.
 The 
{\mbox Bures}
 metric has certainly been the most widely-studied member of this class
\cite{sam,hans,hubner1,hubner2,ditt1,ditt2}.
Two other prominent members are the {\it maximal} \cite{yuenlax} and the {\it Kubo-Mori} ($KM$)
\cite{hasegawa,petz3,michor} 
(also termed Bogoliubov-Kubo-Mori and Chentsov \cite{streater}) 
monotone metrics.

As to the maximal monotone metric, numerical, together with some analytical evidence, strongly indicate that
$V_{max}^{s+n}$ is infinite (unbounded) and that $P_{max}^{s}=0$.
(The supporting 
analytical evidence consists in the fact 
that for the three-dimensional convex set of $2 \times 2$ density matrices,
parameterized by spherical coordinates [$r,\theta,\phi$] in the ``Bloch
ball'', the volume element of the maximal monotone metric is
$r^2 \sin{\theta} (1-r^2)^{-{3/2}}$, the integral of which {\it 
diverges} over
the  ball. Contrastingly, the volume element of the minimal monotone metric is
$r^2 \sin{\theta} (1-r^2)^{-{1/2}}$, the integral over the ball of which is
{\it finite}, namely
$\pi^2$.)

In this analysis, we will seek --- analogously to our study of the
SD/Bures
 metric in \cite{slaterqip} --- to determine $V_{\tilde{KM}}^{s},V_{\tilde{KM}}^{s+n}$ and
thus
their ratio, $P_{KM}^{s}$. (A wiggly line over the acronym for a metric will denote here that we have multiplied that
metric by 4, in order to facilitate comparisons with the 
results of our previous analysis \cite{slaterqip}, which had been presented 
primarily in terms of the SD, rather than the [proportional]
Bures metric. The probabilities themselves --- being ratios --- are, of course, invariant under such a scaling, so the ``wiggle'' is irrelevant
for them.)
In particular, we first 
find compelling numerical evidence that 
\begin{equation} \label{mmm}
V_{\tilde{KM}}^{s+n} = 64 V_{SD}^{s} = \frac{4 \pi^8}{105} \approx 361.468.
\end{equation}
Then, using a  scrambled {\it Faure} sequence \cite{tezuka} composed of
70 million points, rather than the 
scrambled {\it Halton} sequences employed in \cite{slaterqip},
we obtain an estimate that
\begin{equation}
V^s_{\tilde{KM}} \approx  12.6822.
\end{equation}
This leads us to the further estimates that
\begin{equation}
P^s_{KM} \approx  .0350853,
\end{equation}
and
\begin{equation}
\frac{P_{KM}^s}{P_{SD/Bures}^s} \approx .476147.
\end{equation}
Critical to our analysis will be a certain {\it ansatz} that we have
previously employed for similar purposes in \cite{slaterA}.
Contained in the formula for the ``Bures
volume of the set of mixed quantum states'' of Sommers and \.Zyczkowski \cite[eq. (3.18)]{hans}, is the subexpression (following their notation)
\begin{equation} \label{ue}
Q_{N}= \Pi_{\nu < \mu}^{1 \ldots N} \frac{(\rho_{\nu} -\rho_{\mu})^2}{\rho_{\nu} + \rho_{\mu}},
\end{equation}
where $\rho_{\mu},\rho_{\nu}$ ($\mu,\nu=1,\ldots,N$) denote the eigenvalues
of an $N \times N$ density matrix.
The term (\ref{ue}) can equivalently be rewritten using the ``Morozova-Chentsov'' function for the Bures metric \cite[eq. (2.18)]{hans},
\begin{equation} \label{rrr}
 c_{Bures}(\rho_{\mu},\rho_{\nu}) = \frac{2}{\rho_{\nu}
 + \rho_{\mu}}
\end{equation}
as
\begin{equation} \label{xxx}
Q_{N}= \Pi_{\nu < \mu}^{1 \ldots N} (\rho_{\nu}
-\rho_{\mu})^2 c_{Bures}(\rho_{\mu},\rho_{\nu})/2.
\end{equation} 
Our ansatz (working hypothesis) 
is that the replacement of $c_{Bures}$ in the formulas for the Bures volume 
element
 by the particular
Morozova-Chentsov
function corresponding to a given monotone metric ($g$) will
yield the
volume element
 corresponding to $g$. We have been readily able to
validate this for a number of instances
 in the case of the two-level quantum systems  
[$N=2$], using
the general formula for the monotone metrics over 
such systems of Petz and Sud\"ar \cite[eq. (3.17)]{petz1}.

We note here that the Morozova-Chentsov function for the Kubo-Mori metric is \cite[eq. (2.18)]{hans}
\begin{equation} \label{genform}
c_{KM}(\rho_{\mu},\rho_{\nu}) = 
\frac{\log{\rho_{\nu}} -\log{\rho_{\mu}}}{\rho_{\nu} - \rho_{\mu}}.
\end{equation}

To proceed in this study, we first created a MATHEMATICA 
numerical integration program that succeeded to a high
degree of accuracy in reproducing the formula \cite[eq. (4.11)]{hans},
\begin{equation} \label{hall}
C_{N} =  \frac{2^{N^2-N} \Gamma (N^2/2)}{\pi^{N/2} \Gamma (1)  \ldots 
\Gamma (N+1)}
\end{equation} 
for the Hall/Bures
 normalization constants \cite{hall,slaterhall} for various $N$. (These
constants
 form one of the two factors --- along with the volume of the
flag manifold \cite[eqs. (3.22), (3.23)]{hans} --- in determining the total Bures volume.)
Then, in the MATHEMATICA program,
 we replaced the Morozova-Chentsov function for the Bures metric (\ref{rrr}) in the product formula 
(\ref{xxx})
by the one (\ref{genform}) 
for the Kubo-Mori function. For the cases, $N=3,4$ we found
that the new numerical results were to several decimal places of accuracy (and in the case $N=2$, exactly)
 equal to $2^{N(N-1)/2}$ times the comparable result for the Bures
 metric, given by (\ref{hall}). This immediately implies that the $KM$ volumes
of mixed states are also $2^{N (N-1)/2}$ times the corresponding Bures 
volumes (and the same for the $\tilde{KM}$ and SD volumes),
 since the remaining factors involved, that is, the volumes of the flag
manifolds are common to both the Bures and $KM$ cases (as well as to all
the monotone metrics).
Thus, we arrive at our assertion (\ref{mmm}).

We then numerically 
integrated the $\tilde{KM}$
 volume element over a fifteen-dimensional hypercube using points for evaluation in the hypercube determined by scrambled Faure
sequences. (As in \cite{slaterqip}, the fifteen original variables 
parameterizing the $4 \times 4$ density matrices 
were first linearly transformed so as to all lie in the range
[0,1].) These (``low discrepancy'' [computationally intensive]) 
sequences are designed to give a close-to-{\it uniform} coverage
of points over the hypercube, and thus hopefully yield accurate numerical
integration results.

The results for the two monotone metrics considered so far are
presented in Table~\ref{tab:table1}.
\begin{table}
\caption{\label{tab:table1}Estimates based
on the Bures and Kubo-Mori metrics, using a scrambled
Faure sequence of 70 million points for numerical integration. In estimating
$P_{metric}^s$, we use the {\it known} values of $V_{metric}^{s+n}$ given
by (\ref{eq1}) and (\ref{eq2}) --- that is, 5.64794 and 361.468 --- rather 
than the estimates of them reported in the table}
\begin{ruledtabular}
\begin{tabular}{rccc}
metric & $V^{s+n}_{\tilde{metric}}$ & $V^{s}_{\tilde{metric}}$ & 
$P_{metric}^s = V^s/V^{s+n}$ \\
\hline
Bures & 5.64568 & 0.416172 & 0.0736858 \\
Kubo-Mori & 360.757 & 12.6822 &0.0350853 \\
\end{tabular}
\end{ruledtabular}
\end{table}
The tabulated value of $P^s_{SD/Bures} \approx
 0.0736858$ is closer to the conjectured
value of $\frac{8}{11 \pi^2} \approx 0.0736881$ than the estimates
(based on 65 million points of a scrambled Halton sequence)
obtained in \cite{slaterqip}, whether we scale 
there by the true value (\ref{eq1}) 
of $V_{SD}^{s+n}$ --- which gives $ .0737087$ --- or scale by its 
estimated value there --- which gives $ 0.0737012$.
 (In a fully independent analysis based on 70 million
points of a scrambled Halton sequence, we obtained estimates of
$V^{s+n}_{\tilde{KM}}$ of 362.663, of $V^s_{\tilde{KM}}$ of 12.5809 and of
$P^s_{KM}$ of 0.034805.)

Associated with the minimal (Bures) monotone metric is the operator monotone function, $f_{Bures}(t) = (1+t)/2$, and with
the maximal monotone metric, the operator monotone function, $f_{max}(t)=2 t/(1+t)$ \cite[eq. (2.17)]{hans}.
The {\it average}
 of these two functions, that is, $f_{average}(t)=(1+6 t + t^2)/(4 +4 t)$, 
is also necessarily 
operator monotone \cite[eq. (20)]{petz2} and thus 
 yields 
 a monotone metric. Again employing our basic ansatz, we used the associated Morozova-Chentsov function (given by the general formula \cite[p. 2667]{petz1},
$c(x,y)=1/y f(x/y)$)
\begin{equation}
c_{average} = \frac{4 (\rho_{\mu} +\rho_{\nu})}{\rho_{\mu}^2 + 6 \rho_{\mu} \rho_{\nu} + \rho_{\nu}^2},
\end{equation}
in the same quasi-Monte Carlo computations based on the scrambled Faure sequence composed of 70 million points.
We obtained estimates of $V_{average}^{s+n} \approx 1.00888 \cdot 10^{35}$,
$V_{average}^s \approx 2.825 \cdot 10^{22}$, so it appears that their 
ratio,
$P_{average}^s$, is quite close to zero, if not exactly so (as we reasoned
for the case of the maximal monotone metric itself).

Additionally, in an independent set of analyses (Table~\ref{tab:table2}),
 based on 160 million points
of a scrambled Halton sequence, we sought to obtain estimates of the
probability of separability of two arbitrarily coupled qubits based
on three other monotone metrics of interest.
These correspond to the operator monotone functions,
\begin{eqnarray}
f_{WY}(t) = \frac{1}{4} (\sqrt{t}+1)^2; \quad
 f_{GKS}(t)= t^{t/(t-1)}/e; \quad
f_{NI}(t) = \frac{2 (t-1)^2}{(1+t) {\log(t)}^2}.
\end{eqnarray}
The subscript $WY$ denotes the Wigner-Yanase information metric
\cite[sec. 4]{gi} \cite{wy},
 the subscript $GKS$, the Grosse-Krattenthaler-Slater (or ``quasi-Bures'') 
metric
(which yields the common asymptotic minimax and maximin redundancies
for {\it universal} quantum coding \cite[sec. IV.B]{KS} \cite{gillmassar}), and the subscript $NI$,
the ``noninformative'' metric (termed the Morozova-Chentsov metric in 
\cite{slaterclarke}).
\begin{table}
\caption{\label{tab:table2}Estimates based on the
Wigner-Yanase, Grosse-Krattenthaler-Slater and Noninformative monotone
metrics, using a scrambled Halton sequence 
consisting of 160 million points.
In estimating $P_{metric}^s$, as a ratio, 
we are compelled to use the estimated values of $V^{s+n}_{metric}$,
unlike Table I,
since their true values are not presently  known}
\begin{ruledtabular}
\begin{tabular}{rccc}
metric & $V^{s+n}_{\tilde{metric}}$  & $V^{s}_{\tilde{metric}}$ &
 $P^s_{metric} = V^s/V^{s+n}$ \\
\hline
$WY$ & 446.615 & 2.1963 & .0504375 \\
$GKS$ & 166.906 & .9938  & .0610692\\
$NI$ & 3710.31 & 12.616 & .0348745\\
\end{tabular}
\end{ruledtabular}
\end{table}

On a concluding note, let us indicate that the seventy 
million points
of the scrambled {\it Faure} sequence, central to our analyis, were computed
in seven blocks of ten million each, on seven independent/non-communicating
parallel processors (initialized with {\it different} random matrices).
 It is not totally 
clear to us at this stage, whether or not this is likely to be 
inferior to a (more time-demanding) computation on a
{\it single} processor, or to the use of seven processors all initialized with
identical random matrices (cf. \cite{go}). (In any case, it is certain
 that no such question needs to arise for the [fully deterministic] 
scrambled {\it Halton} sequences, also used here and in
\cite{slaterqip}.) This matter is presently under our investigation.
Nevertheless, in any case, we have achieved here rather impressive, 
improved 
convergence to our previously conjectured 
exact value (\ref{fundamental}) of $P^s_{SD/Bures}$.
\begin{acknowledgments}
I wish to express gratitude to the Kavli Institute for Theoretical
Physics for computational support in this research and to Giray \"Okten
for making available his MATHEMATICA program for computing scrambled
Faure sequences.

\end{acknowledgments}

\bibliography{Metrics}

\begin{thebibliography}{38}
\expandafter\ifx\csname natexlab\endcsname\relax\def\natexlab#1{#1}\fi
\expandafter\ifx\csname bibnamefont\endcsname\relax
  \def\bibnamefont#1{#1}\fi
\expandafter\ifx\csname bibfnamefont\endcsname\relax
  \def\bibfnamefont#1{#1}\fi
\expandafter\ifx\csname citenamefont\endcsname\relax
  \def\citenamefont#1{#1}\fi
\expandafter\ifx\csname url\endcsname\relax
  \def\url#1{\texttt{#1}}\fi
\expandafter\ifx\csname urlprefix\endcsname\relax\def\urlprefix{URL }\fi
\providecommand{\bibinfo}[2]{#2}
\providecommand{\eprint}[2][]{\url{#2}}

\bibitem[{\citenamefont{Mkrtchian and Chaltykyan}(1987)}]{vanik}
\bibinfo{author}{\bibfnamefont{V.~E.} \bibnamefont{Mkrtchian}}
  \bibnamefont{and} \bibinfo{author}{\bibfnamefont{V.~O.}
  \bibnamefont{Chaltykyan}}, \bibinfo{journal}{Opt. Commun.}
  \textbf{\bibinfo{volume}{63}}, \bibinfo{pages}{239} (\bibinfo{year}{1987}).

\bibitem[{\citenamefont{Fano}(1983)}]{fano}
\bibinfo{author}{\bibfnamefont{U.}~\bibnamefont{Fano}}, \bibinfo{journal}{Rev.
  Mod. Phys.} \textbf{\bibinfo{volume}{55}}, \bibinfo{pages}{855}
  (\bibinfo{year}{1983}).

\bibitem[{\citenamefont{Braunstein and Caves}(1994)}]{sam}
\bibinfo{author}{\bibfnamefont{S.~L.} \bibnamefont{Braunstein}}
  \bibnamefont{and} \bibinfo{author}{\bibfnamefont{C.~M.} \bibnamefont{Caves}},
  \bibinfo{journal}{Phys. Rev. Lett.} \textbf{\bibinfo{volume}{72}},
  \bibinfo{pages}{3439} (\bibinfo{year}{1994}).

\bibitem[{\citenamefont{Slater}(2002)}]{slaterqip}
\bibinfo{author}{\bibfnamefont{P.~B.} \bibnamefont{Slater}},
  \bibinfo{journal}{Quant. Info. Proc.} \textbf{\bibinfo{volume}{1}},
  \bibinfo{pages}{397} (\bibinfo{year}{2002}).

\bibitem[{\citenamefont{{\.Z}yczkowski
  et~al.}(1998)\citenamefont{{\.Z}yczkowski, Horodecki, Sanpera, and
  Lewenstein}}]{ZHSL}
\bibinfo{author}{\bibfnamefont{K.}~\bibnamefont{{\.Z}yczkowski}},
  \bibinfo{author}{\bibfnamefont{P.}~\bibnamefont{Horodecki}},
  \bibinfo{author}{\bibfnamefont{A.}~\bibnamefont{Sanpera}}, \bibnamefont{and}
  \bibinfo{author}{\bibfnamefont{M.}~\bibnamefont{Lewenstein}},
  \bibinfo{journal}{Phys. Rev. A} \textbf{\bibinfo{volume}{58}},
  \bibinfo{pages}{883} (\bibinfo{year}{1998}).

\bibitem[{\citenamefont{{\.Z}yczkowski}(1999)}]{zycz2}
\bibinfo{author}{\bibfnamefont{K.}~\bibnamefont{{\.Z}yczkowski}},
  \bibinfo{journal}{Phys. Rev. A} \textbf{\bibinfo{volume}{60}},
  \bibinfo{pages}{3496} (\bibinfo{year}{1999}).

\bibitem[{\citenamefont{Slater}(1999{\natexlab{a}})}]{slaterA}
\bibinfo{author}{\bibfnamefont{P.~B.} \bibnamefont{Slater}},
  \bibinfo{journal}{J. Phys. A} \textbf{\bibinfo{volume}{32}},
  \bibinfo{pages}{5261} (\bibinfo{year}{1999}{\natexlab{a}}).

\bibitem[{\citenamefont{Slater}(2000)}]{slaterC}
\bibinfo{author}{\bibfnamefont{P.~B.} \bibnamefont{Slater}},
  \bibinfo{journal}{Euro. Phys. J. B} \textbf{\bibinfo{volume}{17}},
  \bibinfo{pages}{471} (\bibinfo{year}{2000}).

\bibitem[{\citenamefont{Werner}(1989)}]{werner}
\bibinfo{author}{\bibfnamefont{R.~F.} \bibnamefont{Werner}},
  \bibinfo{journal}{Phys. Rev. A} \textbf{\bibinfo{volume}{40}},
  \bibinfo{pages}{4277} (\bibinfo{year}{1989}).

\bibitem[{\citenamefont{Peres}(1996)}]{asher}
\bibinfo{author}{\bibfnamefont{A.}~\bibnamefont{Peres}},
  \bibinfo{journal}{Phys. Rev. Lett.} \textbf{\bibinfo{volume}{77}},
  \bibinfo{pages}{1413} (\bibinfo{year}{1996}).

\bibitem[{\citenamefont{Horodecki et~al.}(1996)\citenamefont{Horodecki,
  Horodecki, and Horodecki}}]{michal}
\bibinfo{author}{\bibfnamefont{M.}~\bibnamefont{Horodecki}},
  \bibinfo{author}{\bibfnamefont{P.}~\bibnamefont{Horodecki}},
  \bibnamefont{and}
  \bibinfo{author}{\bibfnamefont{R.}~\bibnamefont{Horodecki}},
  \bibinfo{journal}{Phys. Lett. A} \textbf{\bibinfo{volume}{223}},
  \bibinfo{pages}{1} (\bibinfo{year}{1996}).

\bibitem[{\citenamefont{Slater}()}]{qq}
\bibinfo{author}{\bibfnamefont{P.~B.} \bibnamefont{Slater}},
  \eprint{quant-ph/0211150}.

\bibitem[{\citenamefont{Sommers and \.Zyczkowski}()}]{hans}
\bibinfo{author}{\bibfnamefont{H.-J.} \bibnamefont{Sommers}} \bibnamefont{and}
  \bibinfo{author}{\bibfnamefont{K.}~\bibnamefont{\.Zyczkowski}},
  \eprint{quant-ph/0304041}.

\bibitem[{\citenamefont{{\"O}kten}(1999)}]{giray1}
\bibinfo{author}{\bibfnamefont{G.}~\bibnamefont{{\"O}kten}},
  \bibinfo{journal}{MATHEMATICA in Educ. Res.} \textbf{\bibinfo{volume}{8}},
  \bibinfo{pages}{52} (\bibinfo{year}{1999}).

\bibitem[{\citenamefont{Sloane and Plouffe}(1995)}]{sloane}
\bibinfo{author}{\bibfnamefont{N.~J.~A.} \bibnamefont{Sloane}}
  \bibnamefont{and} \bibinfo{author}{\bibfnamefont{S.}~\bibnamefont{Plouffe}},
  \emph{\bibinfo{title}{Encyclopaedia of Integer Sequences}}
  (\bibinfo{publisher}{Academic Press}, \bibinfo{year}{1995}).

\bibitem[{\citenamefont{Conway and Guy}(1996)}]{conway}
\bibinfo{author}{\bibfnamefont{J.~H.} \bibnamefont{Conway}} \bibnamefont{and}
  \bibinfo{author}{\bibfnamefont{R.~K.} \bibnamefont{Guy}},
  \emph{\bibinfo{title}{The Book of Numbers}}
  (\bibinfo{publisher}{Springer-Verlag}, \bibinfo{year}{1996}).

\bibitem[{\citenamefont{Petz and Sud\mbox{\"a}r}(1996)}]{petz1}
\bibinfo{author}{\bibfnamefont{D.}~\bibnamefont{Petz}} \bibnamefont{and}
  \bibinfo{author}{\bibfnamefont{C.}~\bibnamefont{Sud\mbox{\"a}r}},
  \bibinfo{journal}{J. Math. Phys.} \textbf{\bibinfo{volume}{37}},
  \bibinfo{pages}{2662} (\bibinfo{year}{1996}).

\bibitem[{\citenamefont{Petz}(1996)}]{petz2}
\bibinfo{author}{\bibfnamefont{D.}~\bibnamefont{Petz}}, \bibinfo{journal}{Lin.
  Alg. Applics.} \textbf{\bibinfo{volume}{244}}, \bibinfo{pages}{81}
  (\bibinfo{year}{1996}).

\bibitem[{\citenamefont{Lesniewski and Ruskai}(1999)}]{lesniewski}
\bibinfo{author}{\bibfnamefont{A.}~\bibnamefont{Lesniewski}} \bibnamefont{and}
  \bibinfo{author}{\bibfnamefont{M.~B.} \bibnamefont{Ruskai}},
  \bibinfo{journal}{J. Math. Phys.} \textbf{\bibinfo{volume}{40}},
  \bibinfo{pages}{5702} (\bibinfo{year}{1999}).

\bibitem[{\citenamefont{Kass}(1997)}]{kass}
\bibinfo{author}{\bibfnamefont{R.~E.} \bibnamefont{Kass}},
  \emph{\bibinfo{title}{Geometrical Foundations of Asymptotic Inference}}
  (\bibinfo{publisher}{John Wiley}, \bibinfo{year}{1997}).

\bibitem[{\citenamefont{H{\"u}bner}(1992)}]{hubner1}
\bibinfo{author}{\bibfnamefont{M.}~\bibnamefont{H{\"u}bner}},
  \bibinfo{journal}{Phys. Lett. A} \textbf{\bibinfo{volume}{63}},
  \bibinfo{pages}{239} (\bibinfo{year}{1992}).

\bibitem[{\citenamefont{H{\"u}bner}(179)}]{hubner2}
\bibinfo{author}{\bibfnamefont{M.}~\bibnamefont{H{\"u}bner}},
  \bibinfo{journal}{Phys. Lett. A} \textbf{\bibinfo{volume}{179}},
  \bibinfo{pages}{226} (\bibinfo{year}{179}).

\bibitem[{\citenamefont{Dittmann}(1999{\natexlab{a}})}]{ditt1}
\bibinfo{author}{\bibfnamefont{J.}~\bibnamefont{Dittmann}},
  \bibinfo{journal}{J. Phys. A} \textbf{\bibinfo{volume}{32}},
  \bibinfo{pages}{2663} (\bibinfo{year}{1999}{\natexlab{a}}).

\bibitem[{\citenamefont{Dittmann}(1999{\natexlab{b}})}]{ditt2}
\bibinfo{author}{\bibfnamefont{J.}~\bibnamefont{Dittmann}},
  \bibinfo{journal}{J. Geom. Phys.} \textbf{\bibinfo{volume}{31}},
  \bibinfo{pages}{16} (\bibinfo{year}{1999}{\natexlab{b}}).

\bibitem[{\citenamefont{Yuen and Lax}(1973)}]{yuenlax}
\bibinfo{author}{\bibfnamefont{H.~P.} \bibnamefont{Yuen}} \bibnamefont{and}
  \bibinfo{author}{\bibfnamefont{M.}~\bibnamefont{Lax}}, \bibinfo{journal}{IEEE
  Trans. Inform. Th.} \textbf{\bibinfo{volume}{19}}, \bibinfo{pages}{740}
  (\bibinfo{year}{1973}).

\bibitem[{\citenamefont{Hasegawa}(1997)}]{hasegawa}
\bibinfo{author}{\bibfnamefont{H.}~\bibnamefont{Hasegawa}},
  \bibinfo{journal}{Rep. Math. Phys.} \textbf{\bibinfo{volume}{39}},
  \bibinfo{pages}{49} (\bibinfo{year}{1997}).

\bibitem[{\citenamefont{Petz}(1994)}]{petz3}
\bibinfo{author}{\bibfnamefont{D.}~\bibnamefont{Petz}}, \bibinfo{journal}{J.
  Math, Phys.} \textbf{\bibinfo{volume}{35}}, \bibinfo{pages}{780}
  (\bibinfo{year}{1994}).

\bibitem[{\citenamefont{Michor et~al.}(2002)\citenamefont{Michor, Petz, and
  Andai}}]{michor}
\bibinfo{author}{\bibfnamefont{P.~W.} \bibnamefont{Michor}},
  \bibinfo{author}{\bibfnamefont{D.}~\bibnamefont{Petz}}, \bibnamefont{and}
  \bibinfo{author}{\bibfnamefont{A.}~\bibnamefont{Andai}},
  \bibinfo{journal}{Infin. Dimens. Anal. Quantum Probab. Relat. Top.}
  \textbf{\bibinfo{volume}{3}}, \bibinfo{pages}{199} (\bibinfo{year}{2002}).

\bibitem[{\citenamefont{Grasselli and Streater}(2001)}]{streater}
\bibinfo{author}{\bibfnamefont{M.~R.} \bibnamefont{Grasselli}}
  \bibnamefont{and} \bibinfo{author}{\bibfnamefont{R.~F.}
  \bibnamefont{Streater}}, \bibinfo{journal}{Infin. Dimens. Anal. Quantum
  Probab. Relat. Top.} \textbf{\bibinfo{volume}{4}}, \bibinfo{pages}{173}
  (\bibinfo{year}{2001}).

\bibitem[{\citenamefont{Faure and Tezuka}(2002)}]{tezuka}
\bibinfo{author}{\bibfnamefont{H.}~\bibnamefont{Faure}} \bibnamefont{and}
  \bibinfo{author}{\bibfnamefont{S.}~\bibnamefont{Tezuka}}, in
  \emph{\bibinfo{booktitle}{Monte Carlo and Quasi-Monte Carlo Methods 2000
  (Hong Kong)}}, edited by \bibinfo{editor}{\bibfnamefont{K.~T.}
  \bibnamefont{Tang}}, \bibinfo{editor}{\bibfnamefont{F.~J.}
  \bibnamefont{Hickernell}}, \bibnamefont{and}
  \bibinfo{editor}{\bibfnamefont{H.}~\bibnamefont{Niederreiter}}
  (\bibinfo{publisher}{Springer}, \bibinfo{year}{2002}), p.
  \bibinfo{pages}{242}.

\bibitem[{\citenamefont{Hall}(1998)}]{hall}
\bibinfo{author}{\bibfnamefont{M.~J.~W.} \bibnamefont{Hall}},
  \bibinfo{journal}{Phys. Lett. A} \textbf{\bibinfo{volume}{242}},
  \bibinfo{pages}{123} (\bibinfo{year}{1998}).

\bibitem[{\citenamefont{Slater}(1999{\natexlab{b}})}]{slaterhall}
\bibinfo{author}{\bibfnamefont{P.~B.} \bibnamefont{Slater}},
  \bibinfo{journal}{J. Phys. A} \textbf{\bibinfo{volume}{32}},
  \bibinfo{pages}{8231} (\bibinfo{year}{1999}{\natexlab{b}}).

\bibitem[{\citenamefont{Gibilisco and Isola}()}]{gi}
\bibinfo{author}{\bibfnamefont{P.}~\bibnamefont{Gibilisco}} \bibnamefont{and}
  \bibinfo{author}{\bibfnamefont{T.}~\bibnamefont{Isola}},
  \eprint{math.PR/0304170}.

\bibitem[{\citenamefont{Wigner and Yanase}(1963)}]{wy}
\bibinfo{author}{\bibfnamefont{E.}~\bibnamefont{Wigner}} \bibnamefont{and}
  \bibinfo{author}{\bibfnamefont{M.}~\bibnamefont{Yanase}},
  \bibinfo{journal}{Proc. Natl. Acad. Sci.} \textbf{\bibinfo{volume}{49}},
  \bibinfo{pages}{910} (\bibinfo{year}{1963}).

\bibitem[{\citenamefont{Krattenthaler and Slater}(2000)}]{KS}
\bibinfo{author}{\bibfnamefont{C.}~\bibnamefont{Krattenthaler}}
  \bibnamefont{and} \bibinfo{author}{\bibfnamefont{P.~B.}
  \bibnamefont{Slater}}, \bibinfo{journal}{IEEE. Trans. Inform. Th.}
  \textbf{\bibinfo{volume}{46}}, \bibinfo{pages}{801} (\bibinfo{year}{2000}).

\bibitem[{\citenamefont{Slater}(2001)}]{gillmassar}
\bibinfo{author}{\bibfnamefont{P.~B.} \bibnamefont{Slater}},
  \bibinfo{journal}{J. Phys. A} \textbf{\bibinfo{volume}{34}},
  \bibinfo{pages}{7029} (\bibinfo{year}{2001}).

\bibitem[{\citenamefont{Slater}(1998)}]{slaterclarke}
\bibinfo{author}{\bibfnamefont{P.~B.} \bibnamefont{Slater}},
  \bibinfo{journal}{Phys. Lett. A} \textbf{\bibinfo{volume}{247}},
  \bibinfo{pages}{1} (\bibinfo{year}{1998}).

\bibitem[{\citenamefont{\mbox{\"O}kten}(2002)}]{go}
\bibinfo{author}{\bibfnamefont{G.}~\bibnamefont{\mbox{\"O}kten}},
  \bibinfo{journal}{Math. Comput. Modell.} \textbf{\bibinfo{volume}{35}},
  \bibinfo{pages}{1221} (\bibinfo{year}{2002}).

\end{thebibliography}

\end{document}